\documentclass[journal]{IEEEtran}

\ifCLASSINFOpdf
\else
   \usepackage[dvips]{graphicx}
\fi
\usepackage{url}

\usepackage{etoolbox}
\robustify\bfseries
\usepackage{graphicx}
\usepackage{amsmath,amssymb,graphicx}
\usepackage[]{algorithm2e}
\usepackage{tabularx}
\usepackage{wasysym}
\usepackage{color,soul}
\usepackage{cite}
\usepackage{balance}
\usepackage{breqn}
\usepackage{makecell}
\usepackage{booktabs}
\usepackage{multirow}
\usepackage{siunitx}
\usepackage{xcolor}
\usepackage{hyperref}

\def\RR{{\mathbb R}}
\usepackage{makecell}

\usepackage{arydshln}

\usepackage[caption=false]{subfig}

\makeatletter
\newcommand*{\bigs}[1]{{\hbox{$\left#1\vbox to9\p@{}\right.\n@space$}}}

\makeatother

\usepackage{xcolor}

\begin{document}

\title{Tackling the Cocktail Fork Problem for Separation and Transcription of Real-World Soundtracks}

\author{Darius Petermann, Gordon Wichern, Aswin Shanmugam Subramanian, Zhong-Qiu Wang, and Jonathan Le Roux
\thanks{
D.\ Petermann is with the Department of Intelligent Systems Engineering, Indiana University, Bloomington, IN 47408, USA (e-mail: daripete@iu.edu).}
\thanks{
G.\ Wichern and J.\ Le Roux are with Mitsubishi Electric Research Laboratories (MERL), Cambridge, MA 02139, USA (e-mail: \{wichern,leroux\}@merl.com).}
\thanks{A.\ S.\ Subramanian was with Mitsubishi Electric Research Laboratories (MERL), Cambridge, MA 02139, USA, and is now with Microsoft Corporation, One Microsoft Way, Redmond, WA 98052, USA (e-mail: asubra13@alumni.jh.edu).}
\thanks{
Z.-Q.\ Wang is with the Language Technologies Institute, Carnegie Mellon University, Pittsburgh, PA 15213, USA (e-mail: wang.zhongqiu41@gmail.com).}
}
\markboth{}
{Shell \MakeLowercase{\textit{et al.}}: Bare Demo of IEEEtran.cls for IEEE Journals}
\maketitle

\begin{abstract}
Emulating the human ability to solve the cocktail party problem, i.e., focus on a source of interest in a complex acoustic scene, is a long standing goal of audio source separation research. Much of this research investigates separating speech from noise, speech from speech, musical instruments from each other, or sound events from each other. In this paper, we focus on the cocktail fork problem, which takes a three-pronged approach to source separation by separating an audio mixture such as a movie soundtrack or podcast into the three broad categories of speech, music, and sound effects (SFX - understood to include ambient noise and natural sound events). We benchmark the performance of several deep learning-based source separation models on this task and evaluate them with respect to simple objective measures such as signal-to-distortion ratio (SDR) as well as objective metrics that better correlate with human perception. Furthermore, we thoroughly evaluate how source separation can influence downstream transcription tasks. First, we investigate the task of activity detection on the three sources as a way to both further improve source separation and perform transcription. We formulate the transcription tasks as speech recognition for speech and audio tagging for music and SFX. We observe that, while the use of source separation estimates improves transcription performance in comparison to the original soundtrack, performance is still sub-optimal due to artifacts introduced by the separation process. Therefore, we thoroughly investigate how remixing of the three separated source stems at various relative levels can reduce artifacts and consequently improve the transcription performance. We find that remixing music and SFX interferences at a target SNR of 17.5 dB reduces speech recognition word error rate, and similar impact from remixing is observed for tagging music and SFX content.
\end{abstract}

\begin{IEEEkeywords}
audio source separation, remixing, speech, music, sound effects, soundtrack, speech recognition, audio tagging, sound event detection
\end{IEEEkeywords}

\IEEEpeerreviewmaketitle

\section{Introduction}

\IEEEPARstart{O}{ver} the last decade and especially with the recent advent of data-driven approaches, many studies have been investigating the separation of audio sources found in media content; whether addressing the separation of speech from non-speech in speech enhancement~\cite{WDL2018, reddy2020interspeech}, speech from other speech in speech separation~\cite{Hershey2016,Drude2019, wichern2019wham}, individual musical instruments in music source separation~\cite{rafii2017musdb, stoter19, manilow2019slakh} or non-speech sound events (or sound effects) in universal sound separation~\cite{kavalerov2019universal,Tzinis_ICASSP2020, pishdadian2020finding,ochiai2020listen}, source separation finds its fit in many fields of application. Separating an audio mixture (e.g., movie soundtrack) into the three broad categories of speech, music, and sound effects (understood to include ambient noise and natural sound events) has however been left largely unexplored despite a wide domain of practical applications. A system properly trained on this task could indeed offer many potential benefits from the consumer standpoint, such as enhancing the listening experience by means of independent volume control over the sources (i.e., remixing), enabling the automatic captioning of sound events, or improving speech transcription accuracy, to name a few.

\begin{figure}[t]
    \centering
        \includegraphics[width=.97\linewidth]{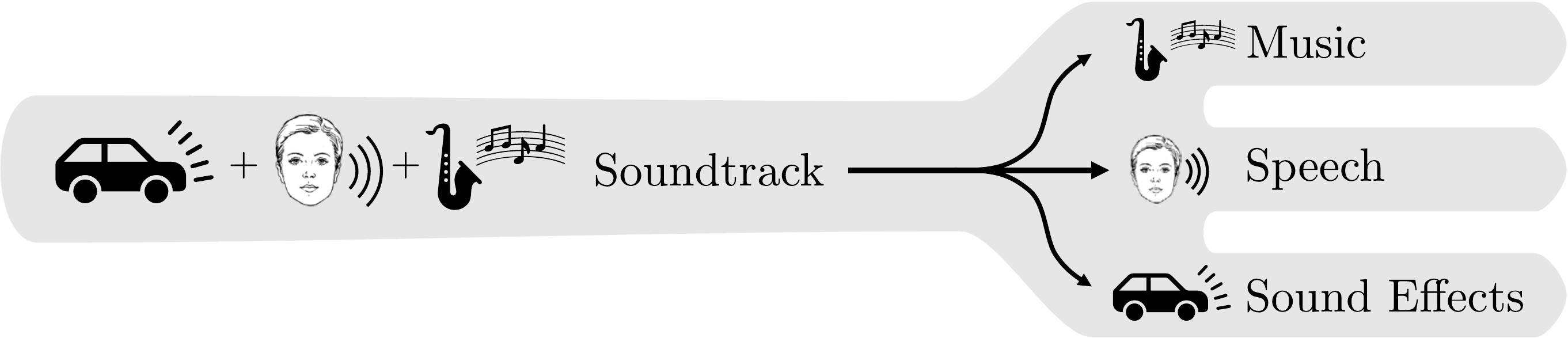}\vspace{-.1cm}
    \caption{Illustration of the cocktail fork problem: given a soundtrack consisting of an audio mixture of speech, music, and sound effects, the goal is to separate it into the three corresponding stems.}\vspace{-.2cm}
    \label{fig:cocktail_fork}
\end{figure}

In our preliminary work %
\cite{petermann2022cfp}, we formalized this new task of music-sound effects (SFX)-speech separation for real-world soundtracks as the cocktail fork problem (CFP), as illustrated in Fig.~\ref{fig:cocktail_fork}, and introduced a dataset specifically tailored towards this task, the Divide and Remaster (DnR) dataset, to foster research on this topic. 
We benchmarked multiple existing popular source separation models and proposed a multi-resolution short-time Fourier transform (STFT) architecture to better address the variety of acoustic characteristics of the three source types. We reported model performance in terms of scale-invariant signal-to-distortion ratio (SI-SDR)~\cite{leroux2019sdr} and showcased the benefit of our proposed model, which produced SI-SDR improvements over the mixture of 11.0 dB for music, 11.2 dB for speech, and 10.8 dB for sound effects. We additionally included an analysis on system performance under different overlapping source conditions.
Finally, we investigated separation performance at different sampling rates.

In this work, we explore the natural extension of the CFP  towards transcription, and explore multiple techniques for integrating source separation and transcription such that the performance of both can be improved. The DnR dataset includes time-stamped annotations for all three source types, in the form of speech transcription for speech, music genre labels for music, and sound-event tags for SFX. 
Specifically, we investigate transcription for the CFP with the following novel contributions:
\begin{itemize}
    \item We present an activity detection system for the three parent classes (music, speech, SFX). By first detecting boundaries, we can then use sophisticated speech recognition and audio tagging models that expect pre-segmented chunks of audio as input. %
    \item We showcase the benefit of integrating the output of our activity detection system with our source separation model by means of various conditioning mechanisms.
    \item We investigate the three classification tasks using well-established pre-trained models. For music genre and sound-event tagging, we evaluate YAMNet, a deep net that predicts 521 audio event classes from the AudioSet-YouTube corpus~\cite{gemmeke2017audioset}. For speech transcription, we evaluate a conformer-based end-to-end automatic speech recognition (ASR) model provided by ESPnet~\cite{li2021espnetse}. The model is pre-trained on LibriSpeech~\cite{librispeech_dataset}, the same dataset DnR is based on.
    \item We explore the idea of source remixing, that is, the act of weighting and adding the separated sources back together, with the goal of increasing classification performance. We show that, in some cases, due to the imperfect nature of the separation, transcription can benefit from source remixing compared to using the raw separation output.
    \item In addition to SDR, we propose to evaluate our source separation models using two other metrics, that are arguably closer to human perception for audio quality assessment: PESQ~\cite{rix2001pesq} for speech and the 2f-model~\cite{kastner20192fmodel} for all three sources.
\end{itemize}
    
In this paper, we aim at further investigating the CFP task, not only from the source separation angle as previously achieved, but from that of transcription as well. 
We hope this paper serves as an important step to a system which would not only be capable of enhancing the listening experience but provide additional semantic understanding as well.

\section{Related Work on the interaction between separation and downstream tasks}
\label{sec:related_work}
The three broad categories of music, speech, and sound effects in audio signals can be found in many different types of settings, from podcasts to radio broadcasts, movies and TV-shows, their presence and overlap are ubiquitous. 
In more elementary scenarios, for example involving only one or two of the three classes, the task of source separation has been well investigated and has shown great promises towards various downstream tasks. For instance, separating individual instruments in musical signals has been found to be beneficial towards music transcription \cite{manilow2020cerberus}, source remixing in the context of music production \cite{woodruff2006remixingSM}, or even towards audio compression \cite{yang2021compression}. 
In the context of noisy speech, speech enhancement (or speech denoising) has shown great promise towards ASR tasks \cite{subramanian2019asr, chang2022chime4}. More recent work proposed a joint audio-tagging and ASR system capable of fulfilling both tasks simultaneously
\cite{moritz2020all}. 
Conversely, several works have considered how classification can improve performance in the context of general sound separation~\cite{Tzinis_ICASSP2020} or music separation \cite{slizovskaia2019class, hung2020multitask}. 
The interaction between voice activity detection and speech enhancement has also been explored  \cite{verteletskaya2010speechactivity, tan2021speech}. 

While the quality of source separation output has no doubt increased dramatically in the deep learning era, artifacts in the separated outputs remain a big problem. For speech enhancement in broadcast applications in particular, recent studies~\cite{westhausen2021reduction, torcoli2021controlling} have shown that remixing the separated speech with some amount of the separated noise can substantially reduce artifacts and improve the listening experience compared to the noisy original. In this paper, we explore how similar remixing ideas can benefit the downstream transcription tasks of ASR, sound event detection, and music genre classification.

Particularly in complex mixtures, where many sounds may interfere with the source of interest, transcription becomes a very challenging task. One way to address this obstacle and potentially improve on the classification output would be to somehow reduce the amount of interfering sources in the mixture and consequently benefit the transcription task. For speech, the idea of using speech enhancement as a front end for recognition has been widely explored \cite{kalinli2010nat, li2021espnetse}. However, these front-end systems are not perfect and may lead to artifacts and unwanted residual noises in the clean speech estimates, which ultimately may negatively affect the downstream ASR task. In \cite{koizumi2021snri}, the authors propose to overcome this limitation by introducing a mechanism controlling an optimal level of noise reduction for the ASR task. An approach for learning whether to use the enhanced speech or the noisy mixture for ASR was presented in~\cite{sato2022learning}, where the authors ultimately found a soft combination between the enhanced signal and the noisy mixture signal performed best. A related study~\cite{iwamoto2022bad} also demonstrated the benefit of remixing the enhanced speech and the noisy mixture for ASR. In this work, we not only evaluate remixing for ASR in the presence of difficult music and sound effect background signals, but also study the benefit of remixing separated signals for sound event tagging and music genre recognition.

In \cite{turpault2020crnn}, the authors evaluated source separation as a pre-processing step to improve the sound-event detection (SED) task by first breaking down mixtures into their constituent sounds. %
SED was then applied by combining the separated sources and the input mixture at different stages in the architecture. Although the separator is not trained jointly in this case (and consequently no active remixing is performed during training), one could argue that the remixing may be done implicitly within the SED network.
Source separation in music applications often focuses on remixing separated musical stems~\cite{yang2022don}. Separation has also been widely used for music transcription (i.e., scoring), either as a pre-processing front-end~\cite{gupta2020automatic, chiu2021source} or within a joint approach %
\cite{manilow2020cerberus,lin2021unified}. %
In \cite{chadna2022framework}, the authors extensively explore the idea of source separation specifically applied to choir ensemble mixtures, allowing for a set of potential downstream applications such as $F_0$ contour analysis, synthesis, transposition, unison analysis, as well as singing group remixing. In \cite{wierstorf2017remix}, the authors investigate existing source separation algorithms and their perceptual impact on songs given various remixing scenarios. The claim suggests that existing separation approaches may suffer from imperfect separation, resulting in perceptible artifacts on individual source estimates, which can then jeopardize downstream task performance. To the best of our knowledge, there has not been any work investigating the task of music source remixing specifically targeted towards downstream labeling tasks such as genre recognition in a manner similar to what we explore in this work.

\section{Methods}
\label{sec:methods}

\subsection{Problem Setup}

\begin{figure*}
    \centering
        \includegraphics[width=1.0\textwidth]{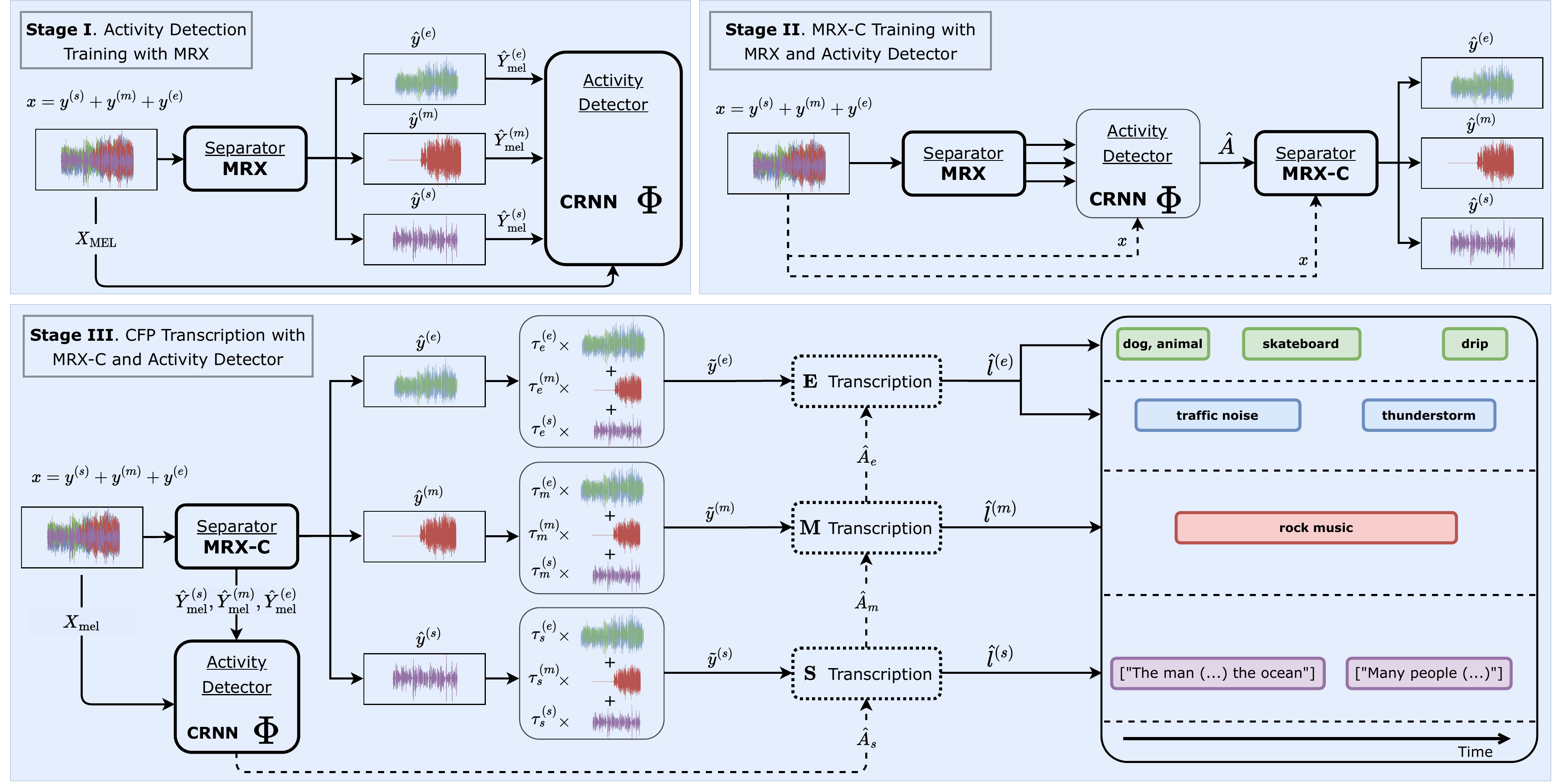}\vspace{-.1cm}
    \caption{Overall separation, activity detection, and transcription pipeline. In \textbf{Stage I}, the activity detector $\Phi_{\text{EI}}$ is trained to provide frame-level activity labels for the three CFP classes (music, speech, SFX). In \textbf{Stage II}, MRX-C, a label-informed version of MRX, is trained using the activity label provided by $\Phi_{\text{EI}}$. Lastly, in \textbf{Stage III}, we proceed to total transcription for a given CFP mixture by first estimating its sources and frame-level labels. The transcription for the three sources is then achieved on the source estimates and using the frame labels.}\vspace{-.6cm}
    \label{fig:overall_pipeline}
\end{figure*}

In this work, we assume that we observe a single-channel mixture $x\in\mathbb{R}^T$  composed of three submixes: 
\begin{equation}
    x = y^{(s)}+y^{(m)}+y^{(e)},
\label{equation:cfp_mixture}
\end{equation}
where $y^{(s)}$ is the submix containing all speech signals, $y^{(m)}$ that of all music signals, and $y^{(e)}$ that of all sound effects. We use the term sound effects (SFX) to broadly cover all sources not categorized as speech or music, and choose it over alternatives such as sound events or noise, as the term is especially relevant to our target application where $x$ is a soundtrack. We here define the cocktail fork problem as that of recovering, from the audio soundtrack $x$, its music, speech, and sound effect submixes, as opposed to extracting individual musical instruments, speakers, or sound effects. 

\begin{table}
\scriptsize
\centering
  \sisetup{table-format=2.1,round-mode=places,round-precision=1,table-number-alignment = center,detect-weight=true}
  \caption{Some examples of events found in the DnR metadata and how we formulate them in the context of the CFP transcription. ``\# classes'' indicates the vocabulary size for each transcription task (e.g., the number of tokens for ASR is 5,000).}
  \vspace{-.3cm}
\begin{tabular}[t]{l c l l l l}
\toprule
Task  & {\# classes} &   $\langle$ {tags,}    & {start\_time,}  &   {end\_time} $\rangle$ \\
\midrule
Music & 11 & $\langle$ {rock} & 2.3 s & 19.5 s $\rangle$ \\
SFX-Fg. & 85 & $\langle$ {dog,bark,animal} & 10.1 s & 13.8 s $\rangle$ \\
SFX-Bg. & 35 & $\langle$ {rain,thunder} & 1.0 s & 30.6 s $\rangle$ \\
Speech & 5,000 & $\langle$ {['the man walks ...']} & 5.9 s & 24.2 s $\rangle$ \\
\bottomrule
\end{tabular}\vspace{-.0cm}
\label{table:transcription_example}
\end{table}

Additionally, we consider the case where the submixes have associated collections $l^{(s)}_{1:N_s}$, $l^{(m)}_{1:N_m}$, $l^{(e)}_{1:N_e}$ 
of metadata labels describing the content for each source type. %
Specifically, as illustrated by the examples in Table~\ref{table:transcription_example}, for speech we consider the speech recognition task where label $l^{(s)}_{i_s}$ represents the transcription of the $i_s$-th utterance and associated time boundaries, with index $i_s\in\{1,\dots,N_s\}$, where $N_s$ is the number of utterances in $y^{(s)}$. Similarly, $l^{(m)}_{i_m}$ represents a music genre label and associated time boundaries for the $i_m$-th music excerpt. For sound effects, $l^{(e)}_{i_e}$ represents a list of audio tags describing the $i_e$-th sound event, along with the associated time boundaries; these tags are further split into two sub-categories of foreground events (SFX-Fg) such as ``dog barking'' and background events (SFX-Bg) such as ``traffic noise,'' which are treated separately for transcription.

In this work, we focus both on separation applications where the goal is to recover estimates $\hat{y}^{(s)}, \hat{y}^{(m)}, \hat{y}^{(e)}$ of the submixes, and on transcription applications where the goal is to estimate metadata label collections $\hat{l}^{(s)}_{1:N_s}$, $\hat{l}^{(m)}_{1:N_m}$, $\hat{l}^{(e)}_{1:N_e}$ given a mixture $x$ and its associated source separation outputs $\hat{y}^{(s)}, \hat{y}^{(m)}, \hat{y}^{(e)}$. 

In the same vein as some of the remixing approaches presented in Section~\ref{sec:related_work}, we explore the idea of source remixing towards the three transcription downstream tasks. 
Our working hypothesis is that, while source separation likely helps the downstream networks to reach a better performance on their respective tasks, it is imperfect, with the presence of added interferences and artifacts, and 
adding back a \emph{down-scaled} version of the predicted constituents in each of the target sources (e.g., music and SFX signals in the speech signal for ASR) prior to performing the transcription may help improve performance. More formally, this can be formulated as follows:
\begin{align}
\Tilde{y}^{(i)} = \tau_i^{(s)} \hat{y}^{(s)} + \tau_i^{(m)} \hat{y}^{(m)} + \tau_i^{(e)} \hat{y}^{(e)}, \quad i \in \{s,m,e\},
 \label{equation:remixing}
\end{align}
where $\Tilde{y}^{(s)}$, $\Tilde{y}^{(m)}$, and $\Tilde{y}^{(e)}$ denote the remixed separated sources, and $\tau_i^{(j)}$ denotes the time invariant gain applied to the separated estimated of source $j$ to obtain the remixed source $i$, e.g., $\tau_m^{(s)}$ is the gain applied to the separated speech estimate $\hat{y}^{(s)}$ when remixing for music transcription. 
Taking the example of speech transcription, the gain $\tau_s^{(s)}$ applied to the speech estimate $\hat{y}^{(s)}$ is set to always remain at unit level, while the gains $\tau_s^{(m)}$ and $\tau_s^{(e)}$ of the interfering sources $\hat{y}^{(m)}$ and $\hat{y}^{(e)}$, respectively, are adjusted (either individually or jointly) to match a target SNR. In the individual case, the gains are set to
\begin{equation}
      \tau_s^{(j)} = \frac{\|\hat{y}^{(s)}\|_2}{\|\hat{y}^{(j)}\|_2} 10^{-
      {\scriptstyle\text{snr}_s^{(j)}}/{20}}, \quad j \in \{m,e\},
\label{equation:snr_individual}
\end{equation}
where $\text{snr}_s^{(m)}$ is the desired SNR of the rescaled speech signal with respect to the rescaled music signal, and $\text{snr}_s^{(e)}$ is defined similarly with respect to the rescaled sound effect signal. Alternatively, we can adjust the gain to reach a desired SNR $\text{snr}_s^{(m+e)}$ of the speech signal with respect to the sum of the two interfering source signals $\hat{y}^{(m+e)}=\hat{y}^{(m)}+\hat{y}^{(e)}$, i.e.,
\begin{equation}
      \tau_s^{(m+e)} = \frac{\|\hat{y}^{(s)}\|_2}{\|\hat{y}^{(m+e)}\|_2} {10^ {-{\scriptstyle\text{snr}_s^{(m+e)}}/{20}}}.
\label{equation:snr_combined}
\end{equation}
Note that, in all cases, we assume that the power of each separated source signal $\hat{y}^{(j)}$ is roughly equal to the power of the corresponding ground-truth source signal ${y}^{(j)}$, which holds in practice if the separation quality is reasonable.
An overview of our proposed approach is shown in Fig.~\ref{fig:overall_pipeline}, and we now describe all stages in detail.

\subsection{Source Separation with MRX}
\label{sec:cfp_ss}
\begin{figure}[t]
\centering
    \includegraphics[width=.95\linewidth]{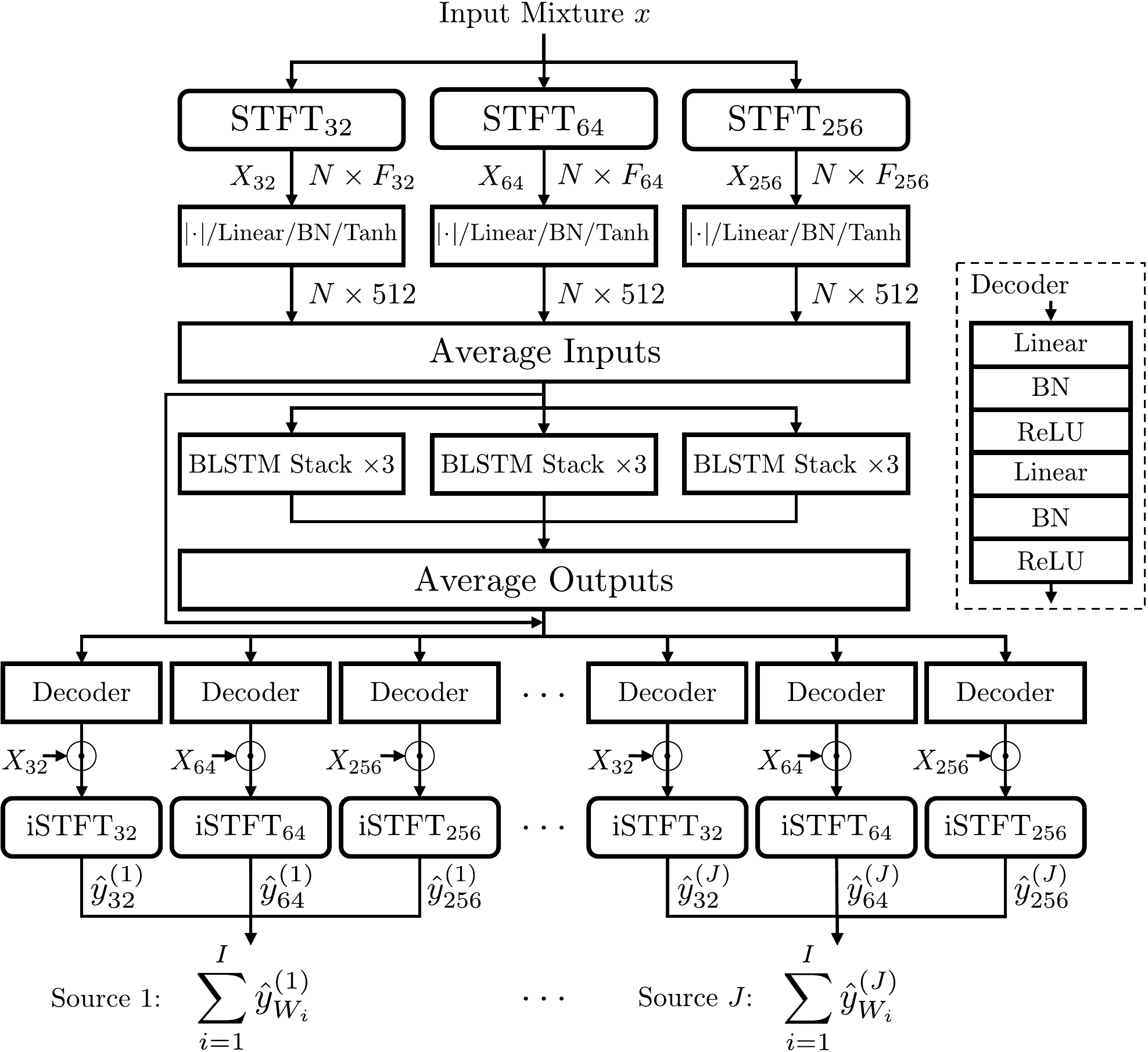}\vspace{-.1cm}
\caption{Multi-resolution CrossNet (MRX) architecture.}\vspace{-.5cm}
\label{fig:xumx_mixed}
\end{figure}

In our preliminary work~\cite{petermann2022cfp}, we found that time-frequency (TF) separation models generally worked well, and we observed additional benefit from jointly evaluating multiple TF resolutions to better handle the diverse acoustic characteristics present in mixtures of speech, music, and sound effects. Therefore, in this work, we use the multi-resolution crossnet (MRX) introduced in \cite{petermann2022cfp} and shown in Fig.~\ref{fig:xumx_mixed} as our main network architecture. MRX takes a time-domain input mixture $x$ and encodes it into $I$ complex spectrograms $X_{W_i}=\text{STFT}_{W_i}(x),$ $i\in\{1,\dots,I\}$ with different STFT resolutions, where $W_i$ denotes the $i$-th window length in milliseconds. Fig.~\ref{fig:xumx_mixed} shows an example with $I=3$ and $\{W_1,W_2,W_3\}=\{32,64,256\}$.

We use the same hop size (e.g., 8 ms in the example of Fig.~\ref{fig:xumx_mixed}) for all resolutions, so they remain synchronized in time, and $N$ denotes the number of STFT frames for all resolutions. In practice, we set the window size in samples to the nearest power of $2$,
and the number of unique frequency bins is denoted as $F_{W_i}$. 
Each resolution is then passed to a fully connected block to convert the magnitude spectrograms of dimension ${N \times F_{W_i}}$ into a consistent
dimension of $512$ across the resolution branches. This allows us to average them together prior to the bidirectional long short-term memory (BLSTM) stacks, whose outputs are averaged once again. 
MRX was inspired by the Cross-Unmix (XUMX) architecture proposed in~\cite{sawata2021all}. However, in our case, the input averaging is intended to allow the network to efficiently combine inputs with multiple resolutions.

The average inputs and outputs of the BLSTM stacks are concatenated and
decoded back into magnitude soft masks $\hat{M}^{(j)}_{W_i}$, one for each of the three sources $j\in\{s,m,e\}$ and each of the $I$ original input resolutions $W_i$. The decoder consists of two stacks of fully-connected layers, each followed by batch normalization (BN) and rectified linear units (ReLU). For a given source $j$, each magnitude mask $\hat{M}^{(j)}_{W_i}$ is multiplied element-wise with the original complex mixture spectrogram  $X_{W_i}$ for the corresponding resolution, a corresponding time-domain signal $\hat{y}^{(j)}_{W_i}$ is obtained via inverse STFT, and the estimated time-domain signal $\hat{y}^{(j)}$ is obtained by summing the time-domain signals at each resolution:
\begin{equation}
\hat{y}^{(j)} = \sum_{i=1}^{I} \hat{y}^{(j)}_{W_i}= \sum_{i=1}^{I}\text{iSTFT}( \hat{M}^{(j)}_{W_i} \odot X_{W_i} ).
\label{equation:mrx}
\end{equation}

For the cocktail fork problem, the network has to estimate a total of $3I$ masks. %
Since ReLU is used as the final mask decoder nonlinearity, the network can freely learn weights for each resolution that best reconstruct the time-domain signal. We use the SI-SDR loss function~\cite{luo2019convTasNet,leroux2019sdr} between the estimated signal $\hat{y}^{(j)}$ and the ground-truth signal $y^{(j)}$ for $j\in\{s,m,e\}$.

We now investigate how source separation and activity detection may benefit from each other.

\subsection{Stage I: Using source separation to improve activity detection} \label{sec:cfp_ad}

As a first step towards transcribing the separated speech, music, and sound effects stems, we focus on detecting the temporal regions where each source is active. Only these detected temporal regions for each source are then fed to the appropriate downstream classifiers, namely, ASR for speech, genre recognition for music, and audio tagging for sound effects. End-to-end (E2E) ASR systems are typically sensitive to extended silent portions occurring between utterances as well as extensive utterance length, and may not work optimally when fed unsegmented signals.  Similarly, for music and sound effects, we can pool over temporal regions to obtain more accurate tagging performance. 

Formally, we define $A=(a_{j,n})_{j,n}\in\{0,1\}^{3\times N}$ as the ground-truth activity labels, with elements $a_{j,n}=1$ indicating that source $j\in\{s,m,e\}$ is active at frame $n$, and $a_{j,n}=0$ that it is inactive. Note that the ground-truth activity labels are defined by the presence of an active excerpt at frame $n$ as defined in the metadata, rather than its energy, so pauses in a speech utterance or a song may still be considered active. 
Our goal is to obtain estimated activity labels $\hat{A}=(\hat{a}_{j,n})_{j,n}\in\mathbb{R}^{3\times N}$, using a neural network $\Phi$ applied to an input representation $R$ derived from the mixture of interest:
\begin{equation}\label{equation:crnn}
    \hat{A}=\Phi(R).
\end{equation}
The network is trained to minimize the binary cross-entropy between $A$ and $\hat{A}$ over the training set:
\begin{multline}
\mathcal{L}(A,\hat{A})=\sum_{j,n}\text{BCE}(a_{j,n},\hat{a}_{j,n})  \\
=\sum_{j,n} \Big(-a_{j,n} \text{log}(\hat{a}_{j,n}) - (1 - a_{j,n}) \text{log}(1 - \hat{a}_{j,n})\Big).
\label{equation:pc_loss}
\end{multline}
At inference time, we apply a median filter over time and threshold $\hat{A}$ to determine boundaries. We denote as $\hat{A}_j$ for the estimated activities for source $j$.

For the core architecture of the neural network $\Phi$, we use a convolutional recurrent neural network (CRNN), as such architectures have proven to be highly effective in the context of sound event detection \cite{pishdadian2020finding, baumann2021crnn, debenito2021multi}, with mel spectrograms as the input representation. The estimated activity labels $\hat{a}_{j,n}$ are in the range $[0,1]$ as they are the outputs of a sigmoid activation function. 
Our baseline $\Phi_{\text{B}}$ simply takes the mel spectrogram $X_{\text{mel}}$ of the mixture as input, and we denote its output as $\hat{A}^{(\text{B,mix})}=\Phi_{\text{B}}(X_{\text{mel}})$.

In \cite{turpault2020crnn}, multiple source-separation-based algorithms are proposed to improve sound event detection. We explore such approaches for the CFP as shown in Stage I of Fig.~\ref{fig:overall_pipeline}. In the first approach, \textbf{early integration} (EI), the input to a CRNN $\Phi_{\text{EI}}$ is formed by stacking the mel spectrograms of the mixture and estimated sources along the channel dimension, i.e.,
\begin{equation}\label{equation:early_integration}
    \hat{A}^{(\text{EI})}=\Phi_{\text{EI}}(\text{stack}(X_{\text{mel}}, \hat{Y}^{(s)}_{\text{mel}}, \hat{Y}^{(m)}_{\text{mel}}, \hat{Y}^{(e)}_{\text{mel}})).
\end{equation}
The input is thus of shape $(4\times N \times F)$ where $F$ is the number of mel bands, and $N$ the number of time frames. In \textbf{middle integration} (MI),  $X_{\text{mel}}$ and the $\hat{Y}^{(j)}_{\text{mel}}$ are each individually input into the CNN block of a CRNN $\Phi_{\text{MI}}$, and their outputs are stacked before being fed to the RNN block. Both CRNNs $\Phi_{\text{EI}}$ and $\Phi_{\text{MI}}$ are trained using Eq.~(\ref{equation:pc_loss}), with the source separation network kept frozen. Finally, in \textbf{late integration} (LI), the baseline CRNN $\Phi_{\text{B}}$ is used to process independently the mel spectrograms $X_{\text{mel}}$ of the mixture and $\hat{Y}^{(j)}_{\text{mel}}$ of each estimated source to obtain the corresponding output probabilities $\hat{A}^{(\text{B,mix})}$ and $\hat{A}^{(\text{B},j)}=\Phi_{\text{B}}(\hat{Y}^{(j)}_{\text{mel}})$ (for example, $\hat{A}^{(\text{B},s)}\in\RR^{3\times N}$ denotes the estimated activity labels of all three sources within the separated speech estimate). These probabilities are then combined to obtain the late integration estimates $\hat{A}^{(\text{LI})}$ as:
\begin{equation}
    \hat{A}^{(\text{LI})} = \frac{1}{2}\hat{A}^{(\text{B,mix})} + \frac{1}{2}\Big(\frac{\hat{A}^{(\text{B},s)}+\hat{A}^{(\text{B},m)}+\hat{A}^{(\text{B},e)}}{3}\Big).
\end{equation}

In this work, all three fusion approaches will be compared in Section \ref{section:results}. We found that the early integration approach was the one that worked best for the CFP activity detection. We refer the interested reader to the original work on the topic \cite{turpault2020crnn} for further technical details.

\subsection{Stage II: MRX-C -- Using activity detection to improve source separation}\label{sec:mrx_c}
As shown in Stage II of Fig.~\ref{fig:overall_pipeline}, we also explore how activity detection can benefit source separation, by conditioning our MRX separation models on the activity detection output. 
In our conditioning approach, which we call MRX-C, we concatenate class probabilities $\hat{A}$ obtained at the output of one of the configurations of Section~\ref{sec:cfp_ad} with the MRX input mixture STFT at each resolution shown in Fig.~\ref{fig:xumx_mixed}. %
Beside concatenation, we also experimented with a FiLM \cite{perez2018film} conditioning approach, which consisted in using a FiLM layer placed after each of the MRX STFT operations in Fig.~\ref{fig:xumx_mixed}. This approach led to very similar results to the concatenation counterpart in preliminary experiments, we thus opted to use the concatenation for all further experiments. 
Because the frame rate of the CRNN outputs may be lower than the STFT frame rate of MRX due to temporal pooling operations, we use nearest neighbor upsampling to match the frame rates.

We further explore the impact of using the ground-truth class labels during the training and inference stages as the upper bound oracle performance. We also explored running two consecutive iterations, or passes, of Stage I and Stage II of Fig.~\ref{fig:overall_pipeline}, which we denote MRX-C$_{\text{2p}}$, where the suffix ``2p'' refers to ``2 passes''. 

As a comparison to the approaches combining activity detection and separation explored in this section, we also consider a multi-task learning approach that jointly does activity detection and source separation~\cite{hung2020multitask, tan2021speech}, referred to as MRX-MTL. Here, an additional decoder layer with three sigmoid outputs is added in parallel to the separation decoders in Fig.~\ref{fig:xumx_mixed} to estimate activity detection. A time average pooling layer with a factor of eight is applied at the input of this decoder to be consistent with the CRNN output resolution, and a weighted binary cross-entropy loss is added to the separation loss for training.

\subsection{Stage III: CFP Transcription}
\label{sec:cfp_transcription}
In this section, we focus on downstream  transcription tasks using the remixed source separation outputs and source activity boundaries. Specifically, we investigate audio tagging for music (in the form of music genre recognition) and sound effects, and ASR for speech. %

\subsubsection{Audio Tagging}
To transcribe music and sound effects, we make use of the powerful pre-trained audio tagging model YAMNet\footnote{\url{https://github.com/tensorflow/models/tree/master/research/audioset/yamnet}}, which predicts 521 audio event tags, including multiple music genres and sound effects classes. YAMNet is based on the MobileNet convolutional architecture~\cite{howard2017mobilenet}, and has been trained on the AudioSet Ontology~\cite{jort201audioset}, a human-labeled corpus derived from short Youtube audio segments. YAMNet operates on 960 ms frames with 50\% overlap, and outputs a vector of  521 class activity probabilities for each frame. The output layer uses sigmoid activation functions, so multiple classes (i.e., tags) can be active for each frame. In practice, we input the entire soundtrack (either the mixture, separated sources, or remixes) into YAMNet, and then average the class probabilities over segments estimated by the activity detector. Example annotations from a soundtrack are shown in Table~\ref{table:transcription_example}. %

For music, we typically do not expect there to be multiple pieces of music playing simultaneously, so we limit the estimated tag to a single one describing the predominant music genre. As we are considering music genre classification, we only take into account the YAMNET outputs corresponding to music genre classes. 

In real-world soundtracks, sound effects typically serve two main purposes: \emph{background events} that usually entail longer and lower amplitude sounds that help set the scene (e.g., rain, traffic), and \emph{foreground events} which are shorter and louder to help tell the story (e.g., gun shot, footsteps). For this reason, we further sub-divide the sound effects audio tagging task into foreground (SFX-Fg) and background (SFX-Bg) sub-tasks as illustrated in Table~\ref{table:transcription_example}. Unlike for music genre, we allow sound events to be labeled with multiple tags from the AudioSet ontology (e.g., dog, bark, animal). Furthermore, sound events from both SFX-Bg and SFX-Fg can overlap in time as shown in the example of Table~\ref{table:transcription_example}. While our goal is to separate and annotate a single sound effects stem, because of the widely different characteristics of SFX-Fg and SFX-Bg sound events, we consider them separately when creating synthetic mixtures in Section~\ref{sec:dataset} and when evaluating audio tagging performance in Section~\ref{subsec:cfp_results_remixing}. However, for evaluating source separation and activity detection performance in Section~\ref{subsec:cfp_results_ss}, we consider SFX-Fg and SFX-Bg jointly as a single sound effects class.

\begin{figure}
    \centering
        \includegraphics[width=.97\linewidth]{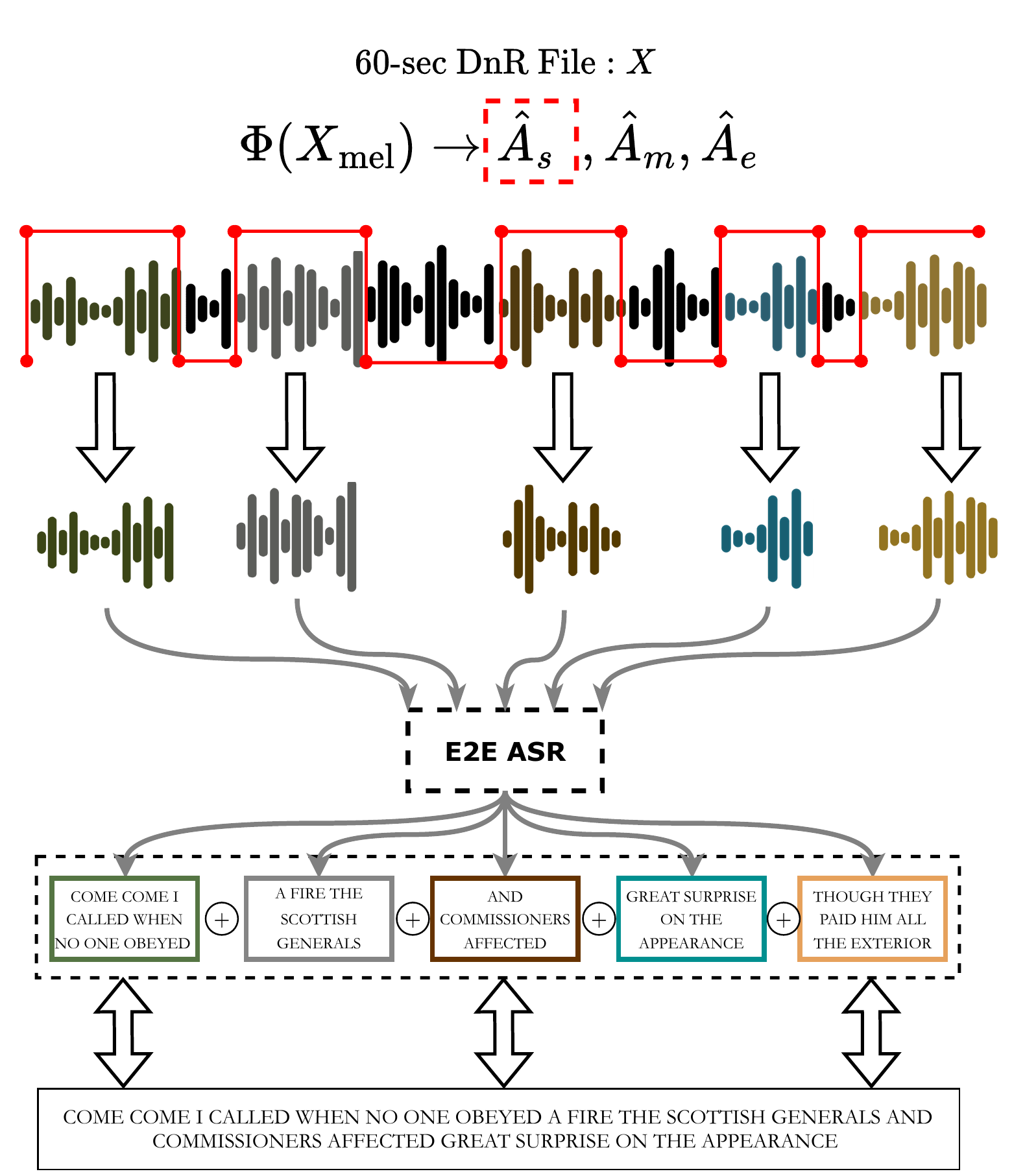}\vspace{-.0cm}
    \caption{Illustration of our ASR evaluation pipeline using ESPnet. We first segment the full 60-second utterance into smaller chunks using $\Phi$'s output. Each of the resulting sub-utterances are then individually fed to ESPnet for decoding. Finally their hypotheses are concatenated and compared against the full-utterance reference.}\vspace{-0.cm}
    \label{fig:espnet_pipeline}
\end{figure}

\subsubsection{Speech Transcription}
We evaluate ASR using a model pre-trained on the LibriSpeech corpus. Specifically, we use a state-of-the-art end-to-end (E2E) model implemented in ESPnet \cite{li2021espnetse} that is based on the Conformer \cite{gulati2020conformer} architecture and uses HuBERT \cite{hsu2021hubert} input features. In general, a soundtrack contains multiple speech utterances interspersed among non-speech regions, which may cause difficulty when feeding the unsegmented soundtrack (be it the mixture, separated sources, or remixes) into an ASR model. Therefore, we evaluate the E2E ASR model using the approach shown in Fig.~\ref{fig:espnet_pipeline}, where the activity detector $\Phi_{\text{EI}}$ first retrieves individual utterances which can then be individually input to the E2E ASR model. The individual utterance hypotheses from the E2E ASR model are then concatenated and evaluated against the full-file reference.

\section{Experimental Setup}
\label{sec:exp_design}

\subsection{Divide and Remaster (DnR) dataset}\label{sec:dataset}
The DnR dataset was introduced in our preliminary work~\cite{petermann2022cfp} with the goal of creating synthetic monophonic soundtracks for training and evaluating source separation algorithms. The dataset is publicly available along with the data creation scripts\footnote{\url{https://cocktail-fork.github.io}}, and further details on the creation process are available in~\cite{petermann2022cfp}. DnR is created by mixing speech from LibriSpeech~\cite{librispeech_dataset}, music from the Free Music Archive (FMA)~\cite{fma_dataset}, and sound effects from the Freesound Dataset 50k (FSD50K) \cite{fsd50k_dataset}. All of the DnR building blocks contain audio at sampling rates of 44.1 kHz or greater (by default LibriSpeech is available at 16 kHz, but the original 44.1 kHz mp3 files are available), so we use 44.1 kHz as the default sampling rate for DnR. This enables real-world listening applications, and can easily be downsampled for transcription where high bandwidth is unnecessary. For FSD50k, we manually classify each of the 200 class labels into one of 3 groups: foreground sounds (e.g., dog bark),  background sounds (e.g., traffic noise), and speech/musical instruments (e.g., guitar, speech).  Speech and musical instrument clips are filtered out to avoid confusion with our speech and music datasets, and we use different mixing rules for foreground and background events.

Many synthetic mixing pipelines for source separation such as wsj0-2mix~\cite{Hershey2016} or speech/music/noise separation~\cite{zhang2021multitask} create fully-overlapped mixtures, with arbitrarily selected source levels. However, models trained on these datasets may not robustly transfer to real-world situations without full-overlap~\cite{menne2019dpcl}. Therefore, we took great care in DnR to make mixtures that are sufficiently realistic in terms of source overlap and relative amplitude level between the three classes. In order to ensure that a mixture could contain multiple full speech utterances and feature a sufficient number of onsets and offsets between the different classes, we decided to make each mixture 60 seconds long.
We do not allow within-class overlap between  speech and music clips, i.e., two music files (or two speech files) will not overlap, but foreground and background sound effects can overlap. This mixing procedure leads to, approximately, $55\%$ of the DnR test set frames having speech, music, and sound effects active, $32\%$  containing two of the three, $10\%$ containing one of the three, and $3\%$ silent frames. 

Regarding the relative amplitude levels across the three classes, after analyzing studies such as~\cite{chaudhuri2018ava} and informal mixing rules from industries such as motion pictures, video games, and podcasting, we follow an approach where speech is generally found at the forefront of the mix, followed by foreground sound effects, then music, and finally background sound effects. 
DnR contains 3,406 60-second mixtures for training, 487 for validation, and 973 for testing.

Along with the audio, DnR also contains transcription metadata in a format similar to the example of Table~\ref{table:transcription_example}. The start and end times correspond to where in the 60 s mixture the original clip was inserted. The gain applied to the clip is also available from the metadata, but is not shown in the example of Table~\ref{table:transcription_example}. For speech utterances, we use the unaltered transcription of the sentence from the LibriSpeech metadata. For music, we list all genres from the FMA annotations, but we use only the top-level genre (corresponding to 16 commonly used labels such as ``jazz'' or ``rock'') in our genre recognition experiments. For sound effects, we list all tags from the FSD50K metadata, and also note whether the clip is used as a foreground or background event.

\subsection{Source Separation}
\label{ssec:separation_design}

\subsubsection{XUMX and MRX models} We consider single-resolution XUMX baselines with various STFT resolutions. We opt to cover a wide range of window lengths $W \in \{32,64,128,256\}$ to assess the impact of resolution on performance. For our proposed MRX model, we use three STFT resolutions of 32, 64, and 256 ms, which we found to work best on the validation set. We use XUMX$_{W}$ to denote a model with a $W$ ms window. We set the hop size to a quarter of the window size. For the MRX model, we determine hop size based on the shortest window. To parse the contributions of the multi-resolution and multi-decoder features of MRX, we also evaluate an architecture adding MRX’s multi-decoder to the best single-resolution model (XUMX$_{64}$), referred to as XUMX$_{64,\text{multi-dec}}$. This results in an architecture of the same size (i.e., same number of parameters) as our proposed MRX model. In all architectures, each BLSTM layer has 256 hidden units and input/output dimension of 512, and the hidden layer in the decoder has dimension 512. For all MRX models, we use a sampling rate of 44.1 kHz.

\subsubsection{Other benchmarks}

We also evaluate our own implementations of Conv-TasNet \cite{luo2019convTasNet} and a temporal convolution network (TCN) with mask inference (MaskTCN). The sampling rate is again 44.1 kHz for all models. MaskTCN uses an identical TCN to the one used internally by
Conv-TasNet, but the learned encoder and decoder are replaced with STFT and iSTFT operations. For MaskTCN, we use an STFT window/hop of 64/16 ms, and for the learned encoder/decoder of Conv-TasNet, we use 500 filters with a window size of 80 samples and a stride of 40. All TCN parameters in both Conv-TasNet and MaskTCN follow the best configuration of~\cite{luo2019convTasNet}. Additionally, we evaluate Open-Unmix (UMX)~\cite{stoter19}, the predecessor to XUMX, by training a separate model for each source, but without the parallel branches and averaging operations introduced by XUMX. 
The Conv-TasNet, UMX, XUMX, and MRX models all use SI-SDR \cite{luo2019convTasNet,leroux2019sdr} as loss function, while MaskTCN uses the waveform domain $L_1$ loss. 

All models are trained on 9 s chunks, except MaskTCN, trained on 6 s chunks, and Conv-TasNet, trained on 2 s chunks; we found these values to lead to best performance under our GPU memory constraints. All models are trained for 300 epochs using ADAM. The learning rate is initialized to $10^{-3}$, and halved if the validation loss is not improved over 3 epochs.

\subsection{Activity Detection}
The activity detection CRNN models described in Section~\ref{sec:cfp_ad} follow the DCASE 2020 Task 4 baseline architecture and its extensions as described in \cite{turpault2020crnn}, based on the publicly available implementation\footnote{\url{https://github.com/turpaultn/dcase20_task4}}.
We use the ADAM optimizer with a learning rate of $10^{-4}$, $\beta_1=0.9$, and $\beta_2=0.999$. We train the systems for 80 epochs and select the weights returning the lowest loss score on the validation set. Prior to training, the data is preprocessed in the following manner. First, we convert all the 60-second long input mixtures into mel spectrograms with 64 bands, using an FFT size of 2048 and hop-length of 512. During training, we randomly sample 10-second long chunks, which translate to 864 temporal frames, from the resulting features. Note that due to the CRNN pooling operations, the number of frames is reduced by a factor of 8 in contrast to the initial input temporal dimension, reducing the number of temporal frames from $864$ at the input down to $108$ at the output. The activity predictions are obtained by applying a median filter of size $5$ over time and thresholding $\hat{A}$ at $0.5$.

\subsection{MRX-C models}

The conditioned MRX-C models described in Section~\ref{sec:mrx_c} follow the same training pipeline as our MRX model described above. For the conditioning portion, the activity labels are obtained for the entirety of the DnR dataset prior to training by using our best performing activity detection model (later described in Section~\ref{subsec:cfp_results_ss}), the early-integration model $\Phi_{\text{EI}}$. During training and inference, the labels are upsampled to the length of the input mixture spectrogram using the nearest neighbor algorithm. %

The multi-task learning model MRX-MTL is optimized on a compound objective function consisting of the SI-SDR and binary cross entropy losses, with their respective weight being $1.0$ and $10.0$.

\subsection{Transcription}

\subsubsection{Music Genre Recognition}

As YAMNet was originally trained on 16 kHz audio data, all audio involved in our YAMNet experiments is first resampled from 44.1 kHz down to 16 kHz. The audio signal goes through a set of transforms prior to network input. First the magnitude spectrogram is obtained through the STFT transform using a window size of 25ms and hop size of 10ms, the spectrogram is then mapped to 64 mel bins covering a 125-7500Hz range in order to obtain its mel transform. Lastly, a stabilized log operation is applied to it. YAMNet uses 960 ms frames overlapped at 50\%. Since the DnR mixtures are 60 s long, we obtain 124 YAMNet frames for each DnR mixture.

One challenge encountered when using YAMNet for the music classification task was the mapping from the 16 top-level genre labels used in the FMA dataset, to one of the 521 sound-event classes from the Audioset ontology used by YAMNet. Ten of the genres had straightforward mapping between FMA and Audioset, namely Pop, Rock, Soul-RnB, Jazz, Country, Electronic, Blues, Hip-hop, Folk, and Classical. The ``International'' genre used by FMA could be mapped to multiple Audioset genres (e.g., ``Music of Asia'', ``Music of Africa''), however, this one-to-many mapping provided very poor performance. The YAMNet ``Vocal Music'' category consistently yielded high confidence when fed music labeled as ``International'' by FMA, as these audio files typically contained singing with little-to-no instrumental parts. Therefore, we mapped ``International'' to ``Vocal Music''. There are 5 FMA genres that we excluded from the classification task as they were overly broad and had no appropriate Audioset genre counterpart (Experimental, Easy Listening, Instrumental, Spoken, and Old-time/historic). Any clips containing these 5 genres were still used for evaluation of music separation and activity detection, but they were ignored when evaluating music genre recognition performance using YAMNet.

\subsubsection{Sound Event Detection}

To filter music and speech clips from FSD50K during the DnR data creation process, we removed a clip if the majority of its tags were related to speech or music. The size of the FSD50K tag vocabulary after this filtering was reduced from 200 to 146. Furthermore, for YAMNet compatibility, we had to exclude 26 additional tags for the followings reasons:
\begin{itemize}
\item The class is an actual parent class in AudioSet, therefore not covered in YAMNet output (e.g., ``Domestic\_sounds\_and\_home\_sounds").
\item The class does not have any examples in AudioSet, therefore it is not covered in YAMNet output (e.g., ``Gull\_and\_seagull").
\item The tag contains a music or speech label (e.g., ``Male\_speech\_and\_man\_speaking"), even though a majority of its tags are not related to music and speech.
\end{itemize}

We then split the remaining 120 tags into SFX-Fg (85 tags) and SFX-Bg (35 tags). We treat the classification tasks for SFX-Fg and SFX-Bg separately, as we expect relatively poor performance for SFX-Bg due to its low relative level and long duration, which likely overlaps with multiple foreground events in DnR.

When using YAMNet outputs for the music genre, SFX-Fg, and SFX-Bg tasks, we first filter the 521 class probabilities output by YAMNet at each time frame to only contain those relevant for the given task (i.e., 11 classes for music, 85 for SFX-Fg, and 35 for SFX-Bg). In practice, we would use the boundaries provided by the activity detector to sum the class probabilities across all relevant frames for a given segment, and then output any tags with a class probability above a threshold. However, in our experiments, whose results are described in Section~\ref{subsec:cfp_results_remixing}, we aim to evaluate how source separation and remixing can aid soundtrack transcription without having activity detection performance play an out-sized role. We therefore use oracle event boundaries for summing YAMNet probabilities. This also allows us to use threshold independent metrics such as mean average (mAP) precision and area under the ROC curve (AUC), without having to account for missed detections and false alarms. Furthermore, it enables evaluation of the SFX-Fg and SFX-Bg tasks separately, given that our activity detector only outputs overall sound effect boundaries.

\subsubsection{Automatic Speech Recognition}
All audio is downsampled to 16 kHz, prior to being input to the HuBERT \cite{hsu2021hubert} feature extraction frontend of the pre-trained Conformer-based \cite{gulati2020conformer} ESPnet model\footnote{\url{https://huggingface.co/espnet/simpleoier_librispeech_asr_train_asr_conformer7_hubert_ll60k_large_raw_en_bpe5000_sp}}. As we will demonstrate in Section~\ref{subsec:cfp_results_remixing}, inputting an entire unsegmented 60 s DnR file, which contains multiple utterances interspersed with noise/silence regions, leads to highly sub-optimal ASR performance. Therefore, using the activity detector to first segment speech regions, and then passing each segment to the ASR model before concatenating all the outputs (as previously discussed and illustrated in Fig.~\ref{fig:espnet_pipeline}) was essential to obtaining acceptable ASR performance. Since we evaluate the concatenated transcriptions with the ground-truth transcription from the entire soundtrack, we can easily compare performance using estimated activity detection boundaries and oracle boundaries in terms of word error rate (WER) and character error rate (CER) using the noisy mixture, separated speech stem, or remixed speech stem.

\section{Experimental Results}
\label{section:results}
\subsection{CFP Separation and activity detection}
\label{subsec:cfp_results_ss}

\begin{table}[t]
\scriptsize
\centering
    \sisetup{
    mode=text, %
    detect-weight, %
    tight-spacing=true,
    round-mode=places,
    round-precision=1,
    table-format=2.1,
    table-number-alignment=center
    }
\caption{
SI-SDR [dB] results of baselines and proposed models on DnR. MUSHRA score predictions from the 2f-model are denoted as $\text{2f-m}$. An extra column is included for PESQ scores on speech.
}\vspace{-.0cm}
\setlength{\tabcolsep}{3pt}
{%
\begin{tabular}[t]{l *{6}{S} S[round-precision=2,table-format=1.2]}
\toprule
&\multicolumn{2}{c}{Music} & \multicolumn{2}{c}{SFX} & \multicolumn{3}{c}{Speech}\\
\cmidrule(lr){2-3} \cmidrule(lr){4-5} \cmidrule(lr){6-8} 
\textbf{Model}              & {SI-SDR} & {2f-m.} & {SI-SDR}  & {2f-m.} & {SI-SDR} &  {2f-m.} &  {PESQ} \\
\midrule
No processing                           & -6.84 & 10.29 & -4.97 & 10.93 & 0.99 & 6.31 & 2.053 \\ 
Oracle PSF~\cite{erdogan2015psf}        & 11.57 & 48.32 & 13.66 & 46.96  & 17.80 & 52.10 & 4.5 \\ 
\midrule
Conv-TasNet~\cite{luo2019convTasNet}           & 0.28 & 14.19  & 1.96 & 13.48  & 8.48  & 19.81 & 2.349 \\
MaskTCN~\cite{luo2019convTasNet} & 1.74 & 20.53 & 3.79 & 21.76 & 9.71  & 27.59 & 2.534 \\

UMX$_{\text{64}}$~\cite{stoter19}              & 3.05 & 22.18 & 4.37 & 21.46 & 11.74  & 30.87 & 2.672  \\

XUMX$_{\text{32}}$~\cite{sawata2021all}   & 2.86 & 21.56 & 4.73 & 21.70 & 11.2  & 29.93 &  2.617  \\
XUMX$_{\text{64}}$~\cite{sawata2021all}   & 3.48 & 22.72 & 5.09 & 22.37 & 11.72  & 30.57 & 2.661 \\
XUMX$_{\text{128}}$~\cite{sawata2021all}  & 3.66 & 24.02 & 5.06 & 23.23 & 11.58  & 30.69 & 2.659  \\
XUMX$_{\text{256}}$~\cite{sawata2021all}  & 2.94 & 22.57 & 4.35 & 22.02 & 10.52  & 29.67 &  2.577 \\

XUMX$_{\text{64,multi-dec}}$              & 3.46 & 23.59 & 5.03 & 22.50 & 11.75  & 31.27 & 2.724 \\
MRX (proposed)                          & 4.21 & 25.64 & 5.69 & 24.67 & 12.33  & 32.85 & 2.846  \\
MRX-C (proposed)                    & \bfseries 4.6 & \bfseries 26.88 & \bfseries 6.1 &  \bfseries 26.13 & \bfseries 12.6 & \bfseries 33.37 & \bfseries 2.874 \\
\bottomrule
\end{tabular}\vspace{-.0cm}
\label{table:pesq_peaq}
}
\end{table}

In this section, we evaluate source separation performance in terms of signal reconstruction metrics for listening applications, as well as the interaction between source separation and sound activity detection.

SI-SDR~\cite{leroux2019sdr} is perhaps the most widely used objective measure for deep learning-based source separation, and was used in our preliminary study~\cite{petermann2022cfp}. However, as shown in~\cite{torcoli2021objective}, SI-SDR is not among the objective metrics most correlated with perceptual quality. Therefore, in this work, we also report results using the {2f-model} \cite{kastner20192fmodel}, which combines two mid-level perceptual features from the Perceptual Evaluation of Audio Quality (PEAQ) \cite{colomes1999peaq} standard (we used the PEAQ implementation from~\cite{kabal2002examination}). The 2f-model was fit to output estimates of MUSHRA scores ranging from 0 to 100, and sound signals were upsampled from 44.1 kHz to 48 kHz for input into the PEAQ model. To avoid any influence due to the scaling of the output, which is particularly an issue for models trained with SI-SDR loss, we normalize all outputs to have the same LUFS as their corresponding ground truth. For the speech source, we also report wideband Perceptual Evaluation of Speech Quality (PESQ) \cite{rix2001pesq}, where sound signals are downsampled to 16 kHz.

Table~\ref{table:pesq_peaq} presents the SI-SDR, PESQ, and 2f values of various models trained and tested on DnR, in addition to the ``No Processing'' condition (lower bound, using the mixture as estimate) and oracle phase sensitive mask~\cite{erdogan2015psf} (a form of upper bound).
For each model, SI-SDR improvements are fairly consistent across source types, despite the differences in their relative levels in the mix, which can be seen in the ``No Processing'' SI-SDR.
In general, we observe that our proposed multi-resolution (MRX) models outperform all single-resolution baselines on all source types in terms of SI-SDR, PESQ, and the PEAQ-based 2f-model scores.
This implies that the network learns to effectively combine information from different STFT resolutions to more accurately reconstruct the target sources.
The performance of $\text{XUMX}_{\text{64,multi-dec}}$ further confirms this hypothesis by performing nearly identically in comparison to $\text{XUMX}_{\text{64}}$, showing that the use of multiple decoders alone does not improve performance.
We also observe that the single-source models (UMX) tend to perform comparably to the cross-source models (XUMX, MRX) for speech, but perform worse for music and SFX. We speculate that because music and SFX are quieter in the mix,
it is harder for the network to isolate them effectively without the support of the other sources, while louder sources (here, speech) do not benefit from joint estimation.

\begin{table}
\scriptsize
\centering
    \sisetup{
    detect-weight, %
    mode=text, %
    tight-spacing=true,
    round-mode=places,
    round-precision=1,
    table-format=2.1,
    table-number-alignment=center
    }
\caption{
SI-SDR [dB] results on the DnR test set. The columns ``Oracle Train'' and ``Oracle Test'' denote whether the model has been trained and tested using oracle or predicted activity labels. $\text{MRX-C}$ refers to the MRX conditioned on activity labels, $\text{MRX-C}_{\text{2p}}$ to MRX conditioned on enhanced activity labels (2\textsuperscript{nd} iterative pass), and MRX-MTL to a multi-task learning version of MRX which performs both SS and activity detection.
}\vspace{-.3cm}
\setlength{\tabcolsep}{4pt}
{%
\begin{tabular}[t]{lcc*{4}{S}}
\toprule
& \multicolumn{2}{c}{Activity labels} & \multicolumn{4}{c}{SI-SDR} \\
\cmidrule(lr){2-3}\cmidrule(lr){4-7}
&{Oracle Train}&{Oracle Test}&{Music}&{Speech}&{SFX}&{Avg.}\\
\midrule
No Processing & \textemdash & \textemdash & -6.8 & 1.0 & -5.0 & -3.6 \\
\midrule
MRX & \textemdash & \textemdash & 4.2 & 12.3 & 5.7 & 7.4 \\
MRX-C & \checkmark & \checkmark & 5.1 & 12.6 & 6.5 & 8.1 \\
MRX-C & \checkmark & $\times$ & 4.4 & 12.5 & 5.9 & 7.6 \\
MRX-C & $\times$ & $\times$ & 4.6 & 12.6 & 6.1 & 7.8\\
MRX-C$_{\text{2p}}$ & $\times$ & $\times$ & 4.6 & 12.6 & 6.1 & 7.8 \\
MRX-MTL & \textemdash & \textemdash & 3.8 & 11.8 & 5.3 & 7.0\\
\bottomrule
\end{tabular}\vspace{-.0cm}
\label{table:model_results}
}
\end{table}

From Table~\ref{table:pesq_peaq}, we see that concatenating activity detection labels with the input spectrogram in the MRX-C model leads to the best overall separation performance, with some gains observed for all metrics compared to MRX. To further evaluate the impact of including activity detection labels as auxiliary inputs for source separation, Table~\ref{table:model_results} compares different settings for activity-conditioned source separation. 
We note that MRX-C leads to an average SI-SDR improvement of 0.7 dB compared to MRX when using oracle information, meaning the upper bound in expected performance improvement is limited, however larger improvements are observed for the quieter and more difficult to separate sources (i.e., music and SFX). When we switch to the realistic setup that does not use oracle information at test time, we see that training using estimated activity detection probabilities leads to a 0.2 dB improvement compared to using oracle information. We also observe that running multiple passes between activity detection and separation (MRX-C$_{\text{2p}}$) led to no performance improvement. Finally, we also considered a multi-task learning setup, where activity detection was used only as an additional training objective (MRX-MTL), but observed a degradation in performance compared to our plain MRX model.

\begin{table}[t]
\scriptsize
\centering
  \sisetup{
    detect-weight, %
    mode=text, %
    round-mode=places,
    round-precision=2,
    table-format=1.2,
    table-number-alignment=center
    }
\caption{Activity detection F-measure for music (M), speech (S), and sound effects (X). The Baseline model uses the mixture as input, MRX-MTL uses multi-task learning for source separation and event detection, while the other rows insert source separation output at different locations inside the network.}\vspace{-.4cm}
{%
\begin{tabular}[t]{l*{6}{S}}
\toprule
& \multicolumn{3}{c}{Event-Based F-Measure} & \multicolumn{3}{c}{Segment-Based F-Measure} \\
\cmidrule(lr){2-4} \cmidrule(lr){5-7} 
 & \multicolumn{1}{c}{M} & \multicolumn{1}{c}{S} & \multicolumn{1}{c}{X} & \multicolumn{1}{c}{M} & \multicolumn{1}{c}{S} & \multicolumn{1}{c}{X} \\
\midrule
Baseline & 0.77 & 0.84 &  0.50 &  0.97 &  0.97 &  0.92   \\
\midrule
MRX-MTL & 0.78 & 0.87 &  0.43 &  0.97 &  0.97 &  0.92   \\
\midrule
Early Integration  & \bfseries 0.82 & 0.88 & \bfseries 0.55 & \bfseries 0.98 & \bfseries 0.98 & \bfseries 0.94 \\
Middle Integration & 0.81 & 0.89 & 0.51 & 0.97 & \bfseries 0.98 & 0.92  \\
Late Integration  & 0.73 & \bfseries 0.91 & 0.50 & 0.97 & \bfseries 0.97 & 0.94 \\
\bottomrule
\end{tabular}\vspace{-.0cm}
\label{table:activity_f_measure}
}
\end{table}
While we have just seen how activity detection can improve separation, we now turn our focus to how separation can aid activity detection. Table~\ref{table:activity_f_measure} displays the sound event detection performance in terms of parent-class F-measure computed using the SED EVAL package~\cite{mesaros2016metrics}. We used a threshold of 0.5, a collar of 750 ms for event-based metrics (with a 20\% offset length), and a time resolution of 1 s for segment-based metrics. Compared to a Baseline taking the mixture signal as input, little improvement is observed using multi-task learning of activity detection and source separation (MRX-MTL). When integrating source separation output into our three-class activity detector, we observed the best performance using early or middle integration, which differs from the permutation-invariant analysis in~\cite{turpault2020crnn} where late integration performed best. We suspect that integrating permutation-invariant source separation outputs at the input or at intermediate layers of the sound event detection network may cause difficulties during training, as the order in which separated outputs are stacked may change between training epochs. However, for problems such as the one studied in this paper, where separated outputs have a fixed ordering, the activity detection network does not have this inconsistency problem and hence early and middle integration perform better than late integration.

\subsection{CFP Transcription and Remixing}
\label{subsec:cfp_results_remixing}

\subsubsection{Audio Tagging (Music and SFX)}
We report audio tagging performance in terms of mAP and mAUC metrics using the implementation from~\cite{chen2020vggsound}. Both mAP and mAUC allow us to evaluate audio tagging performance in a threshold-independent way for situations where multiple tags can be active for a single sound file, and are commonly reported for large-scale audio tagging~\cite{hershey2017cnn, gemmeke2017audioset, fsd50k_dataset}.  Table~\ref{table:remix} displays the audio tagging performance for music, SFX-Fg, and SFX-Bg sources using oracle event boundaries. The ``Noisy'' column is obtained from the original mixture and represents lower-bound performance, and the ``GT'' column represents upper-bound performance obtained using the ground-truth isolated sources. The scores for music are generally higher than for SFX, because there is only a single genre tag for each music segment, the set of possible music labels is smaller than for SFX, and SFX-Fg and SFX-Bg sounds may overlap as discussed in Section~\ref{sec:dataset}. The ``MRX-C'' column from Table~\ref{table:remix} displays audio tagging performance using the separated outputs of MRX-C, our best separation model. It can be seen that for all sources in Table~\ref{table:remix}, source separation improves audio tagging performance compared to the noisy mixture. The improvements for Music and SFX-Fg are larger than those for SFX-Bg, likely because those sources have higher relative levels in the DnR mixes. 

\begin{figure}[htp]
\centering
\subfloat[mAP/mAUC remixing performances for Music]{%
  \includegraphics[width=.97\linewidth]{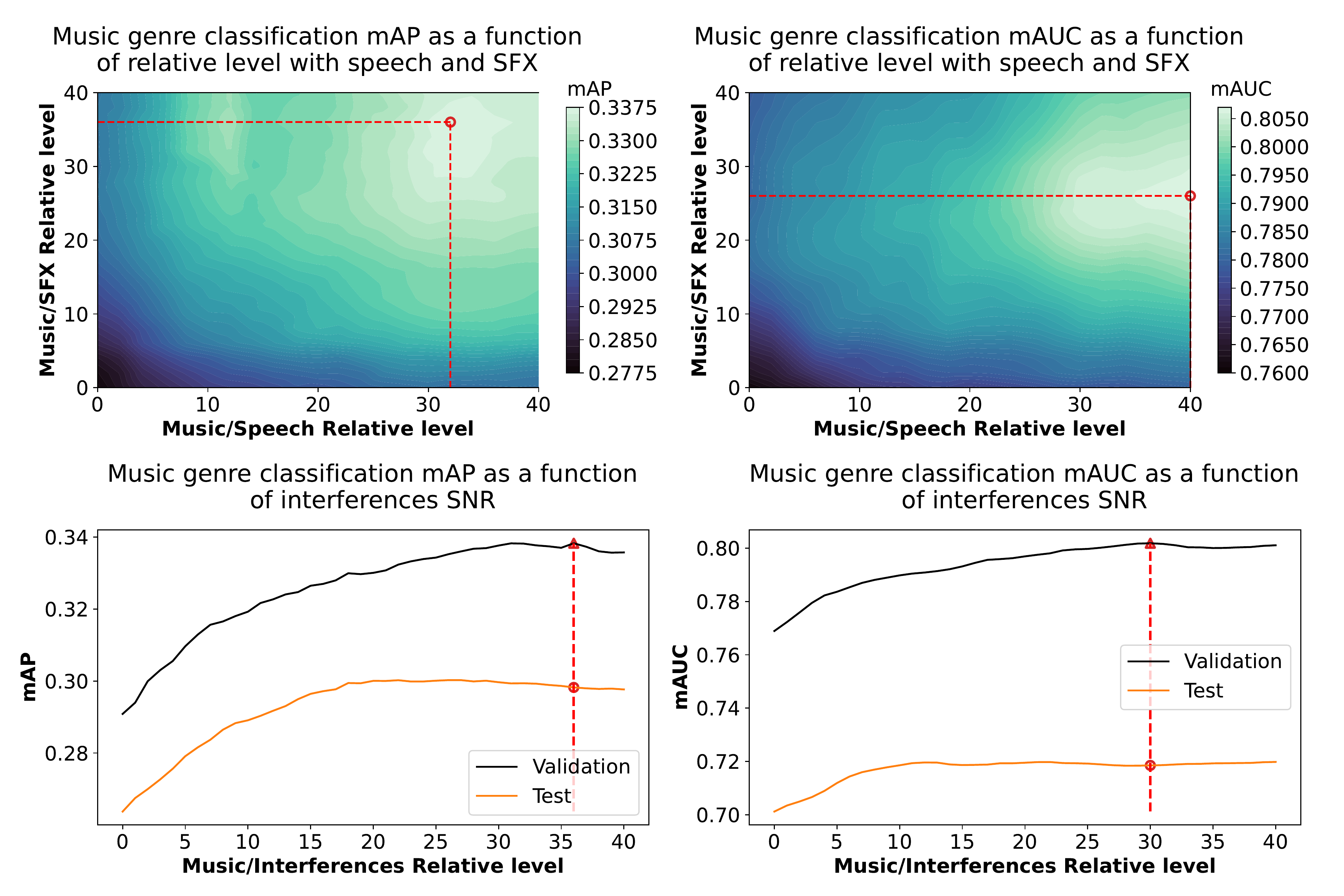}\vspace{-.1cm}
   \label{fig:music_snr}
}

\subfloat[mAP/mAUC remixing performances for SFX-Fg]{%
  \includegraphics[width=.97\linewidth]{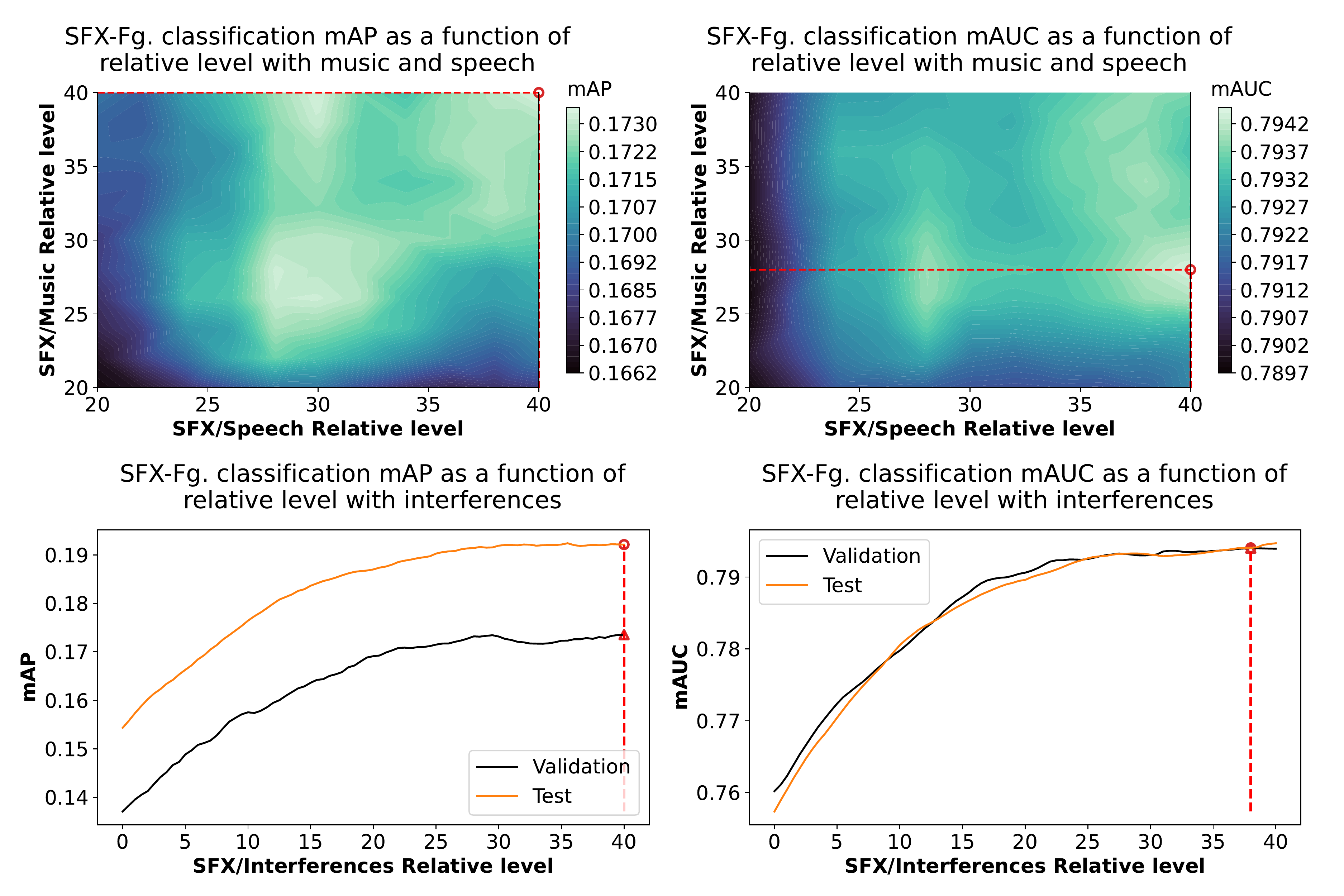}\vspace{-.1cm}
   \label{fig:sfx_fg_snr}
}

\subfloat[mAP/mAUC remixing performances for SFX-Bg]{%
  \includegraphics[width=.97\linewidth]{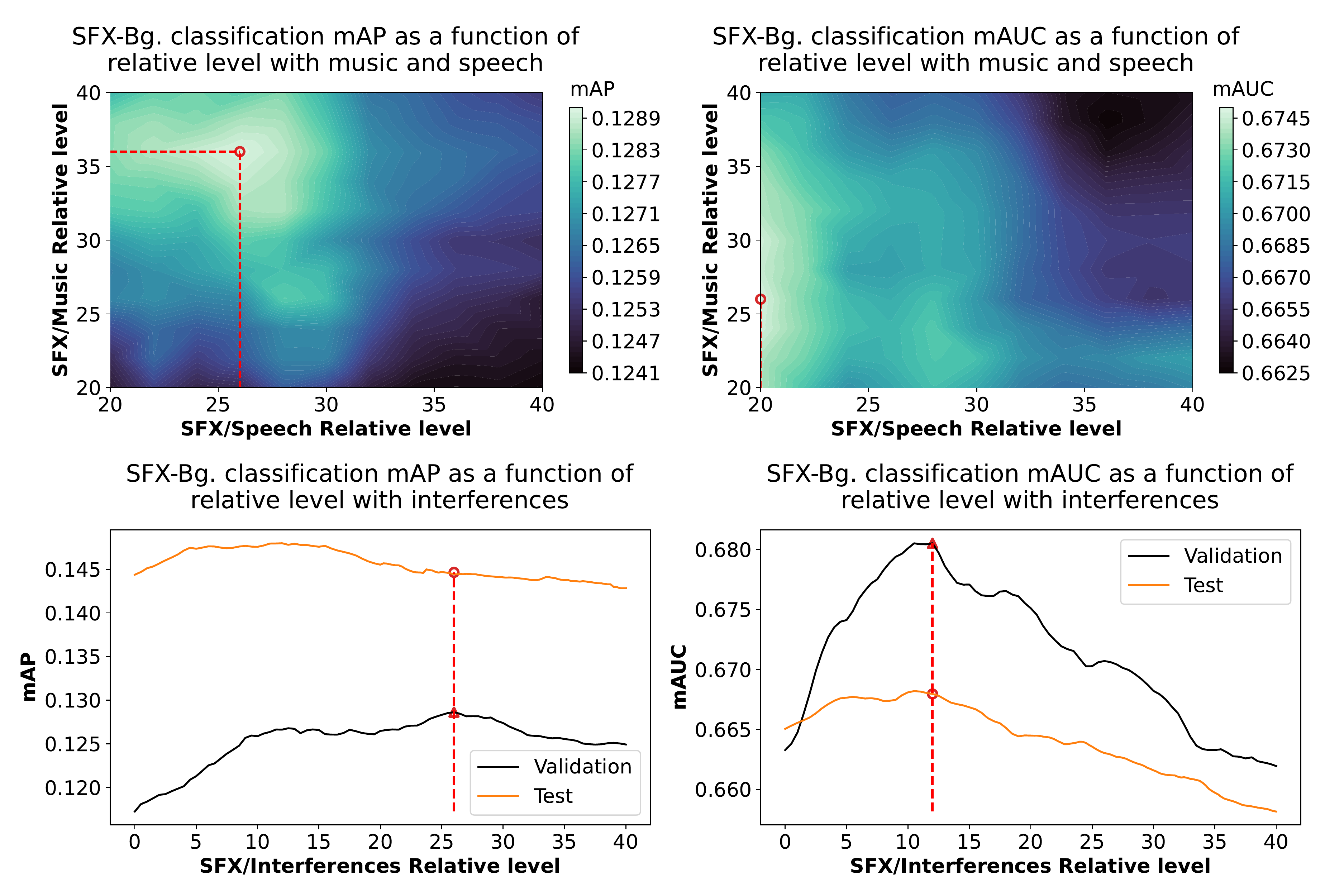}\vspace{-.1cm}
   \label{fig:sfx_bg_snr}
}

\caption{mAUC and mAP music genre classification mappings as a function of Music SNR against interfering sources. The contour plots (first row) show the classification performance on the validation set depending on the SNR against each interfering source individually, while the second row shows the performance on the validation and test sets for the scenario where the two interfering sources are combined in order to compute the SNR. The red symbols along the red dashed lines denote the performance on the test set (circle) where the best validation set performance is observed (triangle).}
\label{fig:results_music_sfx_snr}
\end{figure}

The ``MRX-C Remix'' column in Table~\ref{table:remix} shows the test set results of the systems which obtain the best performance on the validation set for each source in a grid search over remixing gains, as illustrated in Fig.~\ref{fig:results_music_sfx_snr}. %
Figure \ref{fig:results_music_sfx_snr} provides a detailed illustration of the impact of different remixing gains on mAP and mAUC for music and SFX. The first row of Figs.~\ref{fig:music_snr}, \ref{fig:sfx_fg_snr}, and \ref{fig:sfx_bg_snr} shows remixing results on the validation set for the case where interfering source gains are adjusted individually as in~\eqref{equation:snr_individual}, while the second row shows the results on the validation and test sets for the case where the gains are adjusted jointly as in~\eqref{equation:snr_combined}.
For each plot in Fig.~\ref{fig:results_music_sfx_snr}, the data points denoted by the intersections with the red line indicates the best validation set performance for the given metric over different remixing SNR values. In almost all use-cases, we see that both mAP and mAUC benefit from some source remixing in comparison to using the predicted source as the sole signal input. While music and SFX-Fg show only minor classification improvement with remixing, in the case of SFX-Bg, performance peaks at a much lower SNR, meaning that this source specifically benefits by remixing the other sources for its classification task.

\begin{table}[t]
\scriptsize
\centering
  \sisetup{
    mode=text, %
  table-format=1.3,round-mode=places,round-precision=3,table-number-alignment = center,detect-weight=true}
\caption{
mAP and mAUC classification for Music, SFX-Foreground (SFX-Fg.), and SFX-Background (SFX-Bg.). MRX Out. RMX depicts the best remixing use-case scenario.
}\vspace{-.3cm}
\setlength{\tabcolsep}{4.5pt}
{%
\begin{tabular}[t]{l*{8}{S}}
\toprule
&\multicolumn{2}{c}{Noisy} & \multicolumn{2}{c}{MRX-C} & \multicolumn{2}{c}{MRX-C Remix} & \multicolumn{2}{c}{GT}\\
\cmidrule(lr){2-3} \cmidrule(lr){4-5} \cmidrule(lr){6-7} \cmidrule(lr){8-9}
\textbf{Source} & {mAP} & {mAUC} & {mAP} & {mAUC} & {mAP} & {mAUC} & {mAP} & {mAUC}\\
\midrule
Music & 0.233 & 0.682  & 0.297 & 0.723 & 0.3 & 0.718 & 0.336 & 0.772\\
SFX-Fg. & 0.138 & 0.739 & 0.195 & 0.796 & 0.192 & 0.794 & 0.275 & 0.836  \\
SFX-Bg. & 0.137 & 0.653  & 0.143 & 0.656 & 0.148 & 0.666 & 0.258 & 0.767 \\
\bottomrule
\end{tabular}\vspace{-.0cm}
\label{table:remix}
}
\end{table}

\subsubsection{Automatic Speech Recognition}
Table \ref{table:espnet} presents ASR results using ESPnet for three segmentation methods: using the oracle speech utterance onsets/offsets (``Oracle Boundaries''), using the onsets/offsets detected by $\Phi_{\text{EI}}$ (``VAD Boundaries''), and no segmentation (``No Boundaries''). We first note that in the ``No Boundaries'' case, ASR performance degrades dramatically, indicating that our pre-trained ASR model cannot handle 60 s long files containing multiple utterances. We tried training the ASR model from scratch on the DnR dataset speech files, but the performance still significantly lagged that of inputting segmented utterances into the pre-trained ASR model. We also note from Table~\ref{table:espnet} that performance on the original LibriSpeech test set (``Libri.\ Test-Clean'') and the clean DnR speech submix (``DnR Speech GT'') is comparable in the case of oracle boundary segmentation, which is to be expected since they are identical utterances with slightly different levels added during the DnR data creation process. Performance degrades dramatically when the noisy DnR mixture is used for ASR, which is also to be expected since our model is pre-trained using clean speech. When using the separated speech stem from MRX-C, we obtain a 10.5\% absolute (or 60\% relative) reduction in WER compared to the noisy mixture when using VAD boundaries. 

\begin{figure}[htp]
    \centering
        \includegraphics[width=.97\linewidth]{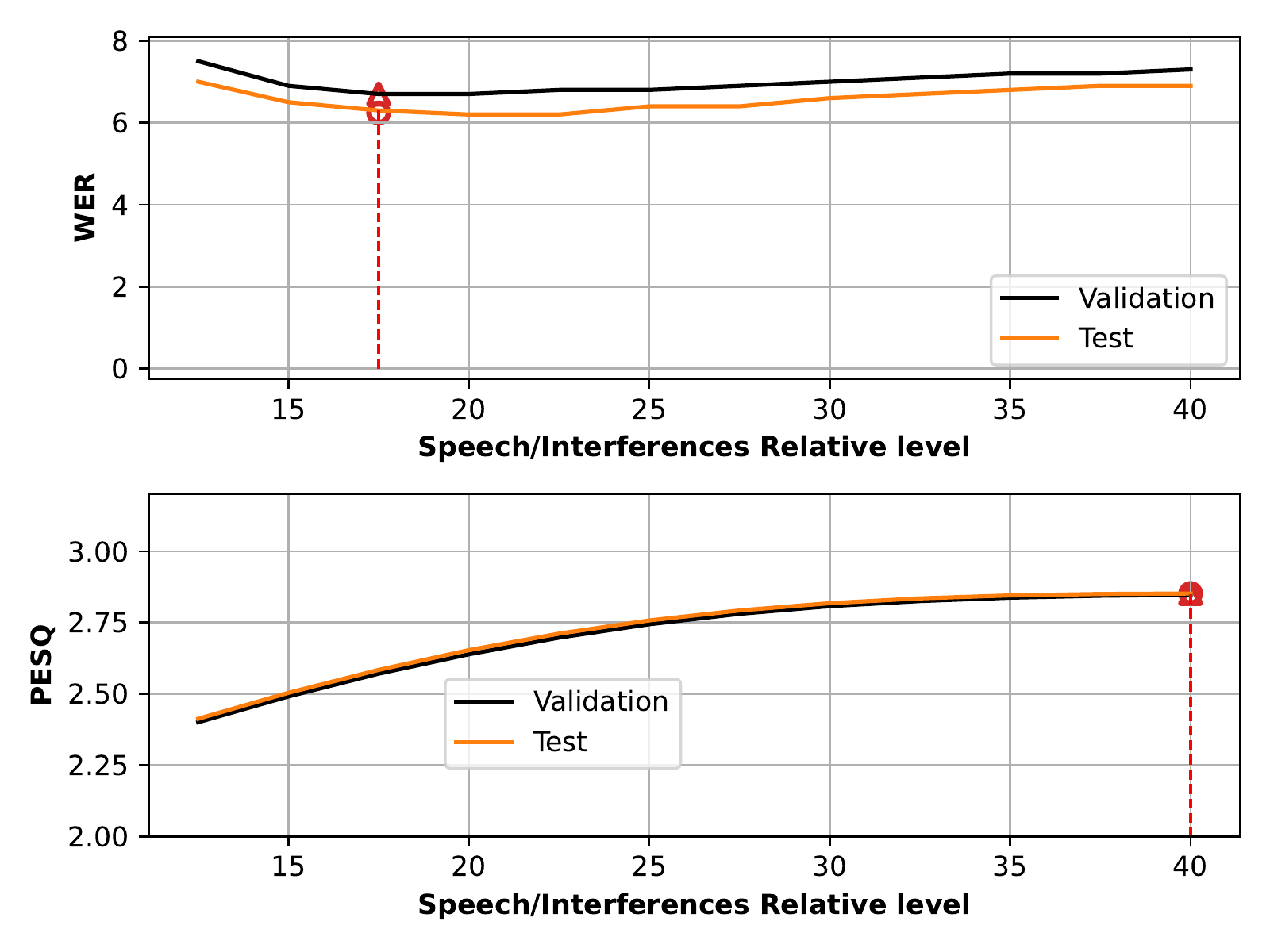}\vspace{-.0cm}
    \caption{WER (\%) and PESQ performances on speech as a function of speech SNR against interfering sources (Music and SFX). Once again the symbols along the red dashed lines denote the points where the best validation set performance is observed (triangle) as well as the test-set performance associated with it (circle).}\vspace{-0.cm}
    \label{fig:pesq_wer}
\end{figure}

In the case of ASR, artifacts introduced by the separation process can degrade performance, and only partially separating the speech has been shown to improve performance~\cite{iwamoto2022bad, sato2022learning, koizumi2021snri}.  Figure~\ref{fig:pesq_wer} displays the performance in terms of WER and PESQ on the validation set obtained for a grid search over the relative level between the remixed separated speech signal and the interference signals. Note that due to the large computational expense of speech decoding, we perform a grid search only for the combined (SFX+music) interference signal as described in~\eqref{equation:snr_combined}. For WER, a minimum is observed at 17.5 dB, which is selected as the gain for the MRX-C Remix system. In the ``MRX-C Remix'' row of Table~\ref{table:espnet}, we observe that by remixing the separated interference signal back with the estimated speech, WER is reduced from 7.0\% down to 6.3\% using VAD boundaries.
In the case of PESQ, the non-remixed speech MRX-C output performs best. That is, MRX-C reports a PESQ score of 2.87, while the best scenario denoted by our remixing experiments is found at SNR set to 40 dB with a PESQ score of 2.85. This underscores the importance of remixing for transcription, while the benefit for listening end-goals is unclear (at least in terms of PESQ), but this is an important topic of future work.

\begin{table}[t]
\scriptsize
\centering
  \sisetup{
  mode=text, %
  table-format=2.1,round-mode=places,round-precision=1,table-number-alignment = center,detect-weight=true,tight-spacing=true}
\caption{
WER and CER (\%) results using different boundary types and input waveforms. The performance on LibriSpeech Test-Clean is also included for reference.
}\vspace{-.3cm}
{%
\begin{tabular}[t]{l*{6}{S}}
\toprule
&\multicolumn{2}{c}{Oracle Boundaries} & \multicolumn{2}{c}{VAD Boundaries} & \multicolumn{2}{c}{No Boundaries}\\
\cmidrule(lr){2-3} \cmidrule(lr){4-5} \cmidrule(lr){6-7}
\textbf{Data} & {WER} & {CER} & {WER} & {CER} & {WER} & {CER}\\
\midrule
Libri.\ Test-Clean & 1.8  & 0.5  & {\textemdash}  & \textemdash  & \textemdash & \textemdash \\
DnR Speech GT & 1.8  & 1.3  & 2.0  & 1.4  & 27.5  & 19.5 \\
Noisy Mixture & 16.4  & 12.3  & 17.5  & 13.0  & 81.7  &  63.7 \\
MRX-C & 6.5  & 4.1  & 7.0  & 4.5  & 36.6  & 25.5  \\
MRX-C Remix & 5.7  & 3.7  & 6.3  & 4.1  & 45.5  & 32.2  \\

\bottomrule
\end{tabular}\vspace{-.0cm}
\label{table:espnet}
}
\end{table}

\section{Conclusions and Discussions}
\label{sec:conclusion}

In this work, we extended our previous work on the cocktail fork problem by tackling transcription for each of the three sources involved: audio tagging for music and sound effect, and ASR for speech. We proposed an activity detection system for the three parent classes and showcased its benefits towards both the separation and transcriptions tasks; we first demonstrated how the system could help improve separation by using the activity labels as conditioning information. Secondly, we described how the integration of an activity detection mechanism was essential in order to tackle real-world soundtrack transcription tasks. We led an extensive investigation to show how source remixing could help towards transcription and demonstrated that mixing back the interfering signals with the isolated source estimates could help improve performance on their associated transcription downstream task. 

In the present work, we explored how transcription benefited from source-remixing strategies, due to the imperfect nature of our separator output. Beside negatively impacting the transcription downstream tasks, the presence of separation artifacts undoubtedly deteriorates the listening experience as well. Moving forward, we aim to explore source-remixing strategies that could minimize perceptual artifacts for the separator output and enhance the listening experience. While we approached this work from a fully supervised angle, for both separation and transcription, taking advantage of the large amount of ``real-world'' unlabeled data available is an important topic for our future work.

\bibliographystyle{IEEEtran}
\bibliography{refs.bib}

\end{document}